\documentclass[twocolumn,showpacs,preprintnumbers]{revtex4-1}
\usepackage{amssymb}
\usepackage{amsfonts}
\usepackage{latexsym}
\usepackage{dcolumn}
\usepackage{amsmath}
\usepackage{graphicx}
\usepackage{psfrag,pstricks}

\begin{document}
\title{Continuous-variable dense coding by optomechanical cavities}
\author{Shabir Barzanjeh$^{1}$}
\author{Stefano Pirandola$^{2}$}
\author{Christian Weedbrook$^{3}$}
\affiliation{$^{1}$Institute for Quantum Information, RWTH Aachen University, 52056 Aachen, Germany}
\affiliation{$^{2}$Department of Computer Science, University of York, York YO10 5GH,
United Kingdom}
\affiliation{$^{3}$Department of Physics, University of Toronto, Toronto, M5S 3G4, Canada}
\date{\today}

\begin{abstract}
In this paper, we show how continuous-variable dense coding can be implemented
using entangled light generated from a membrane-in-the-middle geometry. The
mechanical resonator is assumed to be a high reflectivity membrane hung
inside a high quality factor cavity. We show that the mechanical resonator is
able to generate an amount of entanglement between the optical modes at the
output of the cavity, which is strong enough to approach the capacity of
quantum dense coding at small photon numbers. The suboptimal rate reachable by
our optomechanical protocol is high enough to outperform the classical
capacity of the noiseless quantum channel.

\end{abstract}

\maketitle

\section{Introduction}

The entanglement of quantum states plays an important role in quantum
information~\cite{horo}. Sharing an entangled quantum state, such as an
Einstein-Podolski-Rosen (EPR) state~\cite{chris}, makes it possible to perform
quantum communication processes like quantum dense coding~\cite{bennett},
quantum teleportation~\cite{bennet1}, quantum cryptography~\cite{ekert} and
quantum computational tasks~\cite{nie}. The experimental realizations of these
protocols have been achieved in several physical systems such as photons,
trapped ions, atoms in optical lattices, nuclear magnetic resonance,
etc,~\cite{li,gis}.

A wide range of theoretical and experimental schemes have been also proposed
to generate, observe and/or exploit entanglement using macroscopic
objects~\cite{blen,shab,rip,armour,vitali1,zou,pirs1,pirs2,pirs2003,pirs2004,Mehdip,pat1,pat2,pat3,pat4,pat5,gig}%
. Very recently, it has been shown how mechanical resonators can be used as a
novel tool for generating strong continuous-variable~(CV)
entanglement~\cite{clark}, which may involve optical modes at different
wavelengths~\cite{shab,shab1,len}. Such strong CV\ entanglement can
therefore be exploited to implement quantum information tasks, like dense
coding as studied in this paper.

Quantum dense coding, originally proposed for qubits~\cite{bennett}, provides
a method by which two bits of information can be transmitted by sending only
one qubit, provided that an entangled resource was previously shared by the
parties. This idea was then extended to the CV setting where the
rate at which information is transmitted can potentially be doubled by the use
of EPR states as the source of the entanglement~\cite{bra,ralph,chris}.

In this paper, we show how we can successfully implement the protocol of CV
dense coding by exploiting the optical entanglement at the output of an
optomechanical cavity with a membrane-in-the-middle
geometry~\cite{thompson,jay1}. This system consists of a high finesse cavity
with two fixed-end mirrors and a perfectly-reflecting movable middle mirror,
such as a dielectric membrane. We show the ability of the mechanical resonator
to generate strong entanglement between two output optical beams, in a way
which is robust with respect to the various optomechanical parameters, like
the cavity damping rates, the laser input powers and bandwidths, and the
temperature of the membrane. Then, using this optical entanglement, we prove
that we can perform dense coding with an information rate which closely
approximates the dense coding capacity at small photon numbers, and is also
good enough to outperform the (one-way) classical capacity of the noiseless
quantum channel.

The paper is structured as follows. In Sec.~II, we first give a thorough
theoretical description of the system under consideration and the quantum
Langevin equations (QLEs) are derived and linearized around the semiclassical
steady state. In Sec.~III, we study the steady state of the system and
quantify the entanglement between the output optical modes by using the
logarithmic negativity. In Sec.~IV, we show how the optomechanical source is
able to approach the capacity of dense coding at low energies. Finally, our
conclusions are given in Sec.~V.

\section{Description and dynamics of the optomechanical system}

\begin{figure}[th]
\centering
\includegraphics[width=3.5 in]{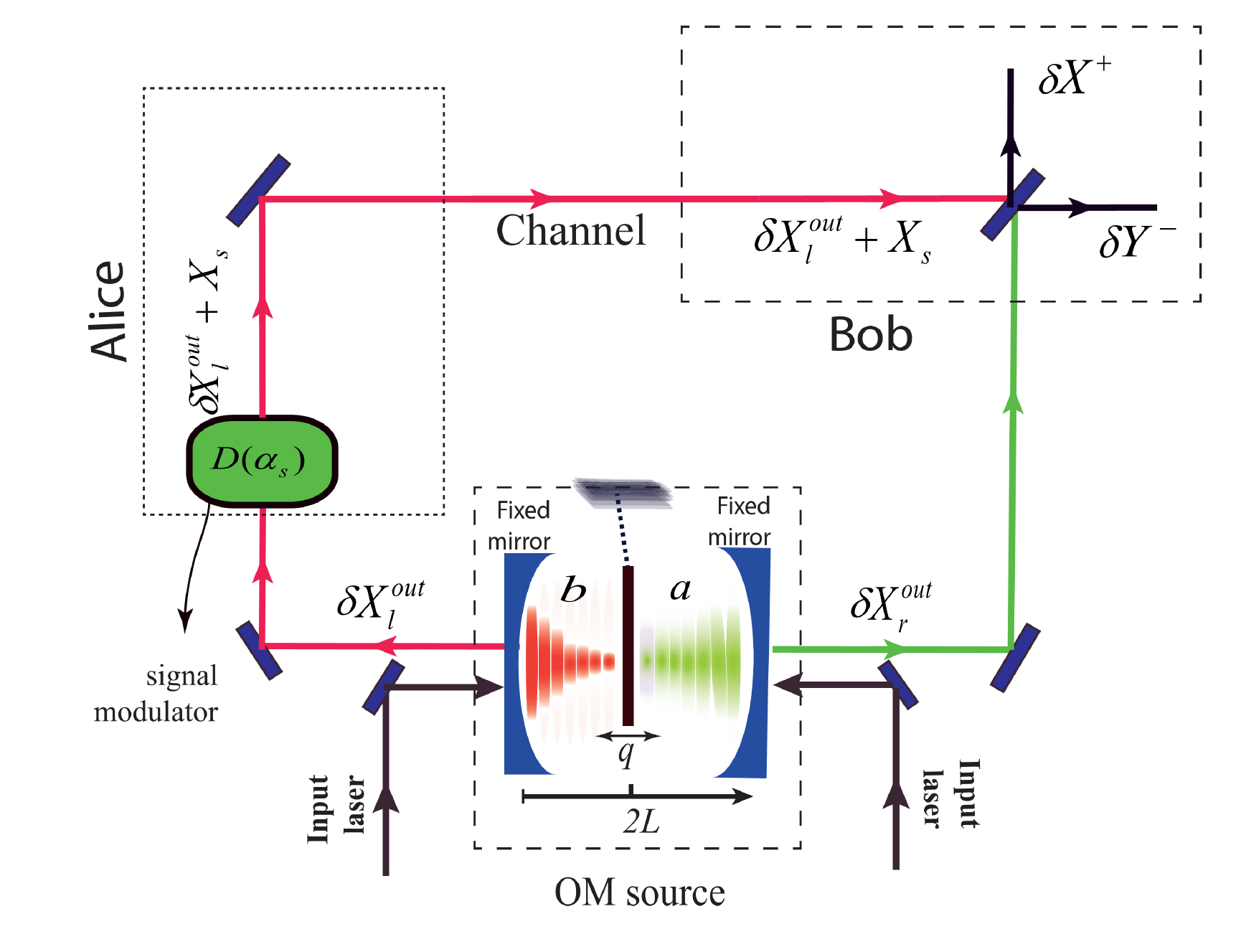}
 \vspace{-1.0cm}
\caption{ (Color online) Protocol of
continuous-variable dense coding equipped with an optomechanical (OM) device as the
source of the entanglement. The preparation of the entanglement consists of a
perfectly reflecting movable middle mirror dividing the cavity into two
separated subcavities right and left with intracavity modes $a$ and $b$,
respectively. The subcavities are excited by two lasers through fixed mirrors.
The left ($l$) and right ($r$) optical outputs of the subcavities are now
entangled and are used to implement continuous-variable dense coding. }%
\label{fig1}%
\end{figure}

We start with a sketch of the system as shown in Fig.~1. A perfectly
reflecting membrane with mass $m$ is placed in the middle of a cavity formed
by two fixed mirrors separated from each other by a distance $2L$. Two strong
coupling fields with amplitudes $\epsilon_{r}$ and $\epsilon_{l}$ and
frequencies $\omega_{0r}$ and $\omega_{0l}$, respectively, are sent into the
cavity through the partially transmitting right and left mirrors. The right
and left subcavities are assumed to be linearly coupled to the displacement of
the membrane with coupling constants $G_{0r}$ and $G_{0l}$, respectively.
Hence the system's time-dependent Hamiltonian takes the form~\cite{bhat}
\begin{align}
H &  =\hbar\omega_{r}a^{\dagger}a+\hbar\omega_{l}b^{\dagger}b+\frac
{\hbar\Omega_{m}}{2}\Big(p^{2}+q^{2}\Big)\label{H1}\\
&  +\hbar\Big(G_{0r}a^{\dagger}a-G_{0l}b^{\dagger}b\Big)q\nonumber\\
&  +i\hbar\epsilon_{r}(a^{\dagger}e^{-i\omega_{0r}t}-ae^{i\omega_{0r}t})+i\hbar\epsilon_{l}(b^{\dagger}e^{-i\omega_{0l}t}-be^{i\omega_{0l}
t}),\nonumber
\end{align}
where $a(b)$ is the annihilation operator for right~(left) subcavity photon
with resonance frequency $\omega_{r}(\omega_{l})$, while $q$ and $p$($[q,p]=i$) are the dimensionless position and momentum operators of the
membrane with frequency $\Omega_{m}$. In Eq.~(\ref{H1}), the optomechanical
coupling constants are expressed by $(i=r,l)$%
\[
G_{0i}=\omega_{i}/L\sqrt{\hbar/m\Omega_{m}}~,
\]
and
\[
\epsilon_{i}=\sqrt{2P_{i}\kappa_{i}/\hbar\omega_{0i}}~,
\]
where $P_{i}$ is the power of the probe lasers impinged inside the cavity
through the right and left mirrors, and $\kappa_{i}$ are the damping rates of
the subcavities' photons via the end mirrors.

In the rotating frame at the frequencies $\omega_{0r}$ and $\omega_{0l}$ of
the driven lasers, we can derive the QLEs for the mirror and the subcavities
variables~\cite{zoller}
\begin{align}
\dot{q}  &  =\Omega_{m}p,\nonumber\\
\dot{p}  &  =-\Omega_{m}q-\gamma_{m}p-G_{0r}a^{\dagger}a+G_{0l}b^{\dagger}b+\xi,\nonumber\\
\dot{a}  &  =-(\kappa_{r}+i\Delta_{0r})a-iG_{0r}qa+\mathrm{\epsilon}_{r}%
+\sqrt{2\kappa_{r}}a_{in},\label{lan1}\\
\dot{b}  &  =-(\kappa_{l}+i\Delta_{0l})b+iG_{0l}qb+\mathrm{\epsilon}_{l}%
+\sqrt{2\kappa_{l}}b_{in},\nonumber
\end{align}
where $\Delta_{0i}=\omega_{i}-\omega_{0i}$ are the detunings, $\gamma_{m}$ is the mechanical damping rate, $a_{in}(t)$ and $b_{in}(t)$ are subcavities' input noises, and $\xi(t)$ is the Brownian noise acting on the mechanical resonator, with correlation function~\cite{pin, law}
\begin{equation}
\langle\xi(t)\xi(t^{\prime})\rangle=\frac{\gamma_{m}}{\Omega_{m}}\int\frac{d\omega}{2\pi}e^{-i\omega(t-t^{\prime})}\omega\Big[\mathrm{coth}%
\big(\frac{\hbar\omega}{2k_{B}T}\big)+1\Big], \label{nois1}
\end{equation}
with $k_{B}$ being the Boltzmann constant, and $T$ the temperature of the reservoir.

In a very high mechanical quality factor regime, i.e., for $Q=\Omega
_{m}/\gamma_{m}\rightarrow\infty$, the mechanical noise is characterized by
white thermal noise~\cite{beng}
\[
\langle\xi(t)\xi(t^{\prime})+\xi(t^{\prime})\xi(t)\rangle/2\simeq\gamma
_{m}(2\bar{n}+1)\delta(t-t^{\prime})~,
\]
with mean excitation number $\bar{n}=[\mathrm{exp}(\hbar\Omega_{m}%
/k_{B}T)-1]^{-1}$. The subcavities' input noises, $a_{in}(t)$ and $b_{in}(t)$,
also obey white-noise correlation functions~\cite{zoller}
\begin{align}
\langle a_{in}(t)a_{in}^{\dagger}(t^{\prime})\rangle &  =\delta(t-t^{\prime
}),\langle a_{in}^{\dagger}(t)a_{in}(t^{\prime})\rangle=0,\label{coropt}\\
\langle b_{in}(t)b_{in}^{\dagger}(t^{\prime})\rangle &  =\delta(t-t^{\prime
}),\langle b_{in}^{\dagger}(t)b_{in}(t^{\prime})\rangle=0,\nonumber
\end{align}
where we have set $\ N(\omega_{i})=[\mathrm{exp}(\hbar\omega_{i}%
/k_{B}T)-1]^{-1}\approx0$, since $\hbar\omega_{i}/k_{B}T\gg1$ at optical frequencies.

The QLEs given in Eq.~(\ref{lan1}) are a set of coupled and nonlinear
differential equations which can be linearized around the semiclassical fixed
points, i.e., $q=q_{s}+\delta q$, $p=p_{s}+\delta p$, $a=\alpha+\delta a$, and
$b=\beta+\delta b$. The fixed points are obtained by setting the time
derivatives to zero, resulting in
\begin{align}
p_{s}  &  =0,\nonumber \\
q_{s}  & =\frac{G_{0l}|\beta|^{2}-G_{0r}|\alpha|^{2}}{\Omega_{m}},\nonumber\\
\alpha & =\frac{\mathrm{\epsilon}_{r}}{\kappa_{r}+i\Delta_{r}},\\
\beta & =\frac{\mathrm{\epsilon}_{l}}{\kappa_{l}+i\Delta_{l}},\nonumber
\end{align}
where $\Delta_{r}=\Delta_{0r}+G_{0r}q_{s}$ and $\Delta_{l}=\Delta_{0l}-G_{0l}q_{s}$ describe the effective detunings of the right and left
subcavities' fields, respectively.

Then, the linear QLEs for the quantum fluctuations of the mirror and the
subcavities' variables are given by%
\begin{align}
\delta\dot{q}  & =\Omega_{m}\delta p,\nonumber\\
\delta\dot{p}  &  =-\Omega_{m}\delta q-\gamma_{m}\delta p+G_{0r}\alpha(\delta
a^{\dagger}+\delta a)\nonumber\\
&  +G_{0l}\beta(\delta b^{\dagger}+\delta b)+\xi,\label{qles2}\\
\delta\dot{a}  &  =-(\kappa_{r}+i\Delta_{r})\delta a-iG_{0r}\alpha\delta
q+\sqrt{2\kappa_{r}}a_{in},\nonumber\\
\delta\dot{b}  &  =-(\kappa_{l}+i\Delta_{l})\delta b+iG_{0l}\beta\delta
q+\sqrt{2\kappa_{l}}b_{in}.\nonumber
\end{align}
where we have chosen the phase references so that
\begin{align}
\alpha=\frac{e^{i\pi}\mathrm{\epsilon}_{r}}{\sqrt{\kappa_{r}+\Delta_{r}}
}~,~\beta=\frac{\mathrm{\epsilon}_{l}}{\sqrt{\kappa_{l}+\Delta_{l}}}~.
\end{align}

\section{Stationary entanglement of the output optical modes}

In this section we study the stationary entanglement between the two optical
modes at the output of the cavity. For this purpose, we derive the stationary
correlation matrix of the system under consideration. First, we rewrite
Eq.~(\ref{qles2}) in terms of the quadrature fluctuactions of the right and
left incavity fields
\begin{align*}
\delta X_{r}  &  =\frac{\delta a+\delta a^{\dagger}}{\sqrt{2}}~,~\delta
Y_{r}=\frac{\delta a-\delta a^{\dagger}}{i\sqrt{2}}~,\\
\delta X_{l}  &  =\frac{\delta b+\delta b^{\dagger}}{\sqrt{2}}~,~\delta
Y_{l}=\frac{\delta b-\delta b^{\dagger}}{i\sqrt{2}}~,
\end{align*}
and the corresponding input noise operators
\begin{align*}
X_{r}^{in}  &  =\frac{ a_{in}+ a_{in}^{\dagger}}{\sqrt{2}
}~,~ Y_{r}^{in}=\frac{a_{in}- a_{in}^{\dagger}}{i\sqrt{2}
}~,\\
X_{l}^{in}  &  =\frac{ b_{in}+ b_{in}^{\dagger}}{\sqrt{2}
}~,~ Y_{l}^{in}=\frac{ b_{in}- b_{in}^{\dagger}}{i\sqrt{2}
}~.
\end{align*}
Thus, we have
\begin{align}
\delta\dot{q}  &  =\Omega_{m}\delta p,\nonumber\\
\delta\dot{p}  &  =-\Omega_{m}\delta q-\gamma_{m}\delta p+G_{r}\delta X_{r}
+G_{l}\delta X_{l}+\xi,\nonumber\\
\delta\dot{X_{r}}  &  =-\kappa_{r}\delta X_{r}+\Delta_{r}\delta Y_{r}
+\sqrt{2\kappa_{r}}X_{r}^{in},\nonumber\\
\delta\dot{Y_{r}}  &  =-\kappa_{r}\delta Y_{r}-\Delta_{r}\delta X_{r}
+G_{r}\delta q+\sqrt{2\kappa_{r}}Y_{r}^{in},\label{qles3}\\
\delta\dot{X_{l}}  &  =-\kappa_{l}\delta X_{l}+\Delta_{l}\delta Y_{l}
+\sqrt{2\kappa_{l}}X_{l}^{in},\nonumber\\
\delta\dot{Y_{l}}  &  =-\kappa_{l}\delta Y_{l}-\Delta_{l}\delta X_{l}%
+G_{l}\delta q+\sqrt{2\kappa_{l}}Y_{l}^{in},\nonumber
\end{align}
with effective optomechanical coupling constants
\[
G_{i}=\frac{2\omega_{i}}{L}\sqrt{\frac{P_{i}\kappa_{i}}{m\Omega_{m}\omega
_{0i}(\kappa_{i}^{2}+\Delta_{i}^{2})}}~~(i=r,l)~.
\]

Now Eq.~(\ref{qles3}) can be written in the compact form
\begin{equation}
\mathbf{\dot{u}}(t)=\mathbf{Au}(t)+\mathbf{n}(t)~, \label{eqQ}
\end{equation}
by introducing the vectors
\begin{align*}
\mathbf{u}(t)  &  =[\delta q(t),\delta p(t),\delta X_{l}(t),\delta
Y_{l}(t),\delta X_{r}(t),\delta Y_{r}(t)]^{T},\\
\mathbf{n}(t)  &  =[0,\xi(t),\sqrt{2\kappa_{l}}X_{l}^{in},\sqrt{2\kappa_{l}%
}Y_{l}^{in},\sqrt{2\kappa_{r}}X_{r}^{in},\sqrt{2\kappa_{r}}Y_{r}^{in}]^{T},
\end{align*}
and defining the drift matrix
\begin{equation}
\mathbf{A}=\left(
{\begin{array}{*{20}c} 0 & {\Omega _m } & 0 & 0 & 0 & 0 \\ { - \Omega _m } & { - \gamma _m } & {G_l } & 0 & {G_r } & 0 \\ 0 & 0 & { - \kappa _l } & {\Delta _l } & 0 & 0 \\ {G_l } & 0 & { - \Delta_r } & { - \kappa _l } & 0 & 0 \\ 0 & 0 & 0 & 0 & { - \kappa _r } & {\Delta _r } \\ {G_r } & 0 & 0 & 0 & { - \Delta _r } & { - \kappa _r } \\ \end{array}}%
\right)  ~. \label{driftA}%
\end{equation}

Solving Eq.~(\ref{eqQ}) we can derive the evolution of the quadrature vector
$\mathbf{u}(t)$. In turn, this solution provides the evolution of the output
quadrature vector%
\begin{align*}
&  \mathbf{u}^{out}(t)\\
&  =[\delta q(t),\delta p(t),\delta X_{l}^{out}(t),\delta Y_{l}^{out}
(t),\delta X_{r}^{out}(t),\delta Y_{r}^{out}(t)]^{T},
\end{align*}
describing the mechanical resonator and the optical fields at the output of
the two subcavities. In particular, we are interested in the stationary
covariance matrix (CM) $\mathbf{V}^{out}$ describing the asymptotic
correlations between the previous modes. This CM has generic element%
\begin{equation}
V_{ij}^{out}=\underset{t\rightarrow\infty}{\lim}\frac{1}{2}\left\langle
u_{i}^{out}(t)u_{j}^{out}(t)+u_{j}^{out}(t)u_{i}^{out}(t)\right\rangle ~,
\label{cor1}
\end{equation}
where $\left\langle ...\right\rangle $ denotes the average on the stationary
state of the system.

In the frequency domain, the stationary CM takes the form~\cite{geni}
\begin{align}
\mathbf{V}^{out}=\int &  d\omega\boldsymbol{\Upsilon}(\omega
)\Big(\mathbf{\tilde{M}}^{ext}(\omega)+\mathbf{P}_{out}%
\Big)\nonumber\label{vmat}\\
&  \times\mathbf{D}_{ext}\Big(\mathbf{\tilde{M}}^{ext}(\omega)^{\dagger
}+\mathbf{P}_{out}\Big)\boldsymbol{\Upsilon}^{\dagger}(\omega),
\end{align}
where $\mathbf{\tilde{M}}^{ext}(\omega)=(i\omega+A)^{-1}$,
\begin{align*}
\mathbf{P}_{out}  &  =\mathrm{diag}[0,0,1/2k_{l},1/2k_{l},1/2k_{r}
,1/2k_{r}]~,\\
\mathbf{D}^{ext}  &  =\mathrm{diag}[0,\gamma_{m}(2\bar{n}_{b}+1),2\kappa_{l},2\kappa_{l},2\kappa_{r},2\kappa_{r}]~,
\end{align*}
and $\boldsymbol{\Upsilon}(\omega)$ is the Fourier transform of
\[
\boldsymbol{\Upsilon}(t)=\left(
{\begin{array}{*{20}c} {\delta (t)} & 0 & 0 & 0 & 0 & 0 \\ 0 & {\delta (t)} & 0 & 0 & 0 & 0 \\ 0 & 0 & R_l & -I_l & 0 & 0 \\ 0 & 0 & I_l & R_l & 0 & 0 \\ 0 & 0 & 0 & 0 & R_r & -I_r \\ 0 & 0 & 0 & 0 & I_r & R_r \\ \end{array}}%
\right)  ~,
\]
where $R_{j}=\sqrt{2\kappa_{j}}\mathrm{Re}[g_{j}(t)]$ and $I_{j}=\sqrt{2\kappa_{j}}\mathrm{Im}[g_{j}(t)]\,\,(j=r,l)$are determined by the causal filter
functions~\cite{van,geni} $g_{j}(t)$, with bandwidths $1/\tau_{j}$ and central
frequencies $\Omega_{j}$. \begin{figure}[th]
\centering
\includegraphics[width=3.2in]{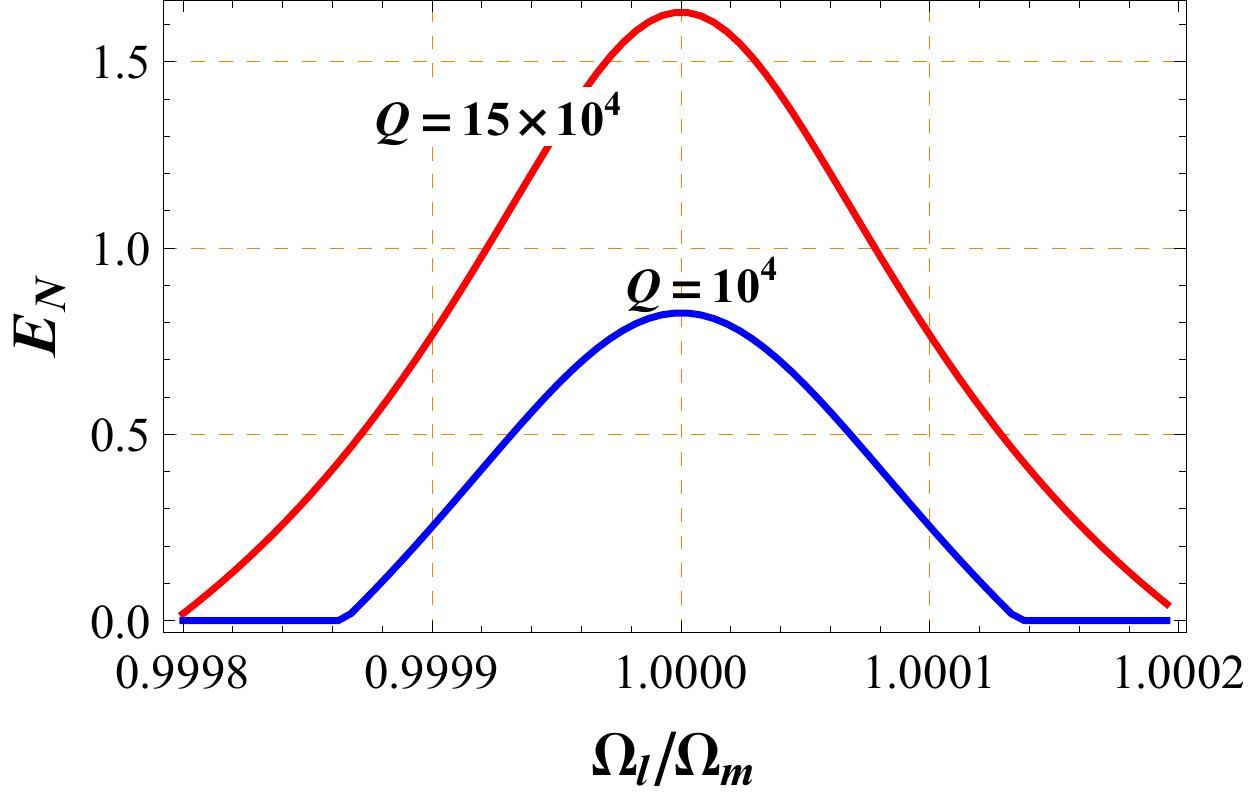}\caption{ (Color online) Logarithmic
negativity $E_{N}$ between the optical output modes of the two subcavities
versus the normalized frequency $\Omega_{l}/\Omega_{m}$ for two different
values of the mirror's quality factor $Q$ at a fixed temperature $T=1$ K with
$\Omega_{r}=-\Omega_{m}$. The subcavities detuning have been fixed at
$\Delta_{r}=-\Omega_{m}$, $\Delta_{l}=\Omega_{m}$ while the other parameters
are $\Omega_{m}/2\pi=10$~MHz, $\kappa_{r}=0.4\Omega_{m}$, $P_{r}=10$~mW, $L=1$
mm, $\kappa_{l}=0.1\Omega_{m}$, $P_{l}=48~$mW, $m=10~$ng.}%
\label{fig2}%
\end{figure}

From the global CM of Eq.~(\ref{vmat}) we extract the reduced CM
$\mathbf{V}^{\prime}$\ of the output optical modes with quadrature
fluctuations $\delta X_{l}^{out},\delta Y_{l}^{out},\delta X_{r}^{out}$ and
$\delta Y_{r}^{out}$. This matrix can be written in the blockform%

\begin{equation}
\mathbf{V}^{\prime}=\left(
\begin{array}
[c]{cc}%
L\mathbf{I} & \mathbf{C}\\
\mathbf{C}^{T} & R\mathbf{I}%
\end{array}
\right)  ,~\mathbf{C}=\left(
\begin{array}[c]{cc}
-C & C^{\prime}\\
C^{\prime} & C
\end{array}
\right)  ~, \label{loga1}
\end{equation}
where $L,R\geq1/2$, $C\geq0$ and $C^{\prime}$ is numerically small compared
with the other matrix elements. This CM completely characterizes the
stationary Gaussian state of the output cavity modes. In particular, this CM
approximates that of an EPR state with cross correlations of the kind $\delta
X_{l}^{out}\approx-\delta X_{r}^{out}$ and $\delta Y_{l}^{out}\approx\delta
Y_{r}^{out}$.

In order to study the conditions under which the output optical modes are
entangled, we consider the logarithmic negativity $E_{N}$~\cite{vid} 
given by%
\begin{equation}
E_{N}=\mathrm{max}[0,-\mathrm{ln}(2\zeta)]~, \label{loga}
\end{equation}
where $\zeta$ is the least partially-transposed symplectic eigenvalue of
$\mathbf{V}^{\prime}$~\cite{chris,Alex}. This is given by
\begin{equation}
\zeta=\sqrt{\frac{\Lambda(\mathbf{V}^{\prime})-\sqrt{\Lambda(\mathbf{V}^{\prime})^{2}-4\mathrm{det}\mathbf{V}^{\prime}}}{2}}~,
\end{equation}
with $\Lambda(\mathbf{V}^{\prime})=L^{2}+R^{2}-2\mathrm{det}\mathbf{C}$.

\begin{figure}[th]
\centering
\includegraphics[width=3.6in]{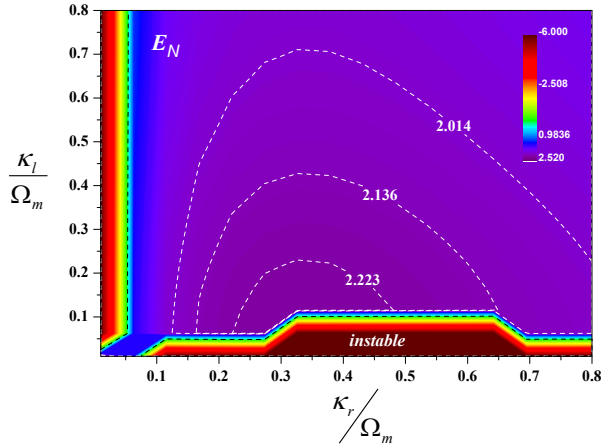}
 \vspace{-1.0cm}
\caption{ (Color online) Logarithmic
negativity $E_{N}$ between the optical output modes of the two subcavities
versus the normalized damping rates $\kappa_{l}/\Omega_{m}$ and $\kappa
_{r}/\Omega_{m}$ at fixed temperature $T=1$ K and $\Omega_{r}=-\Omega
_{l}=-\Omega_{m}$. Other parameters are the same as those in Fig.~\ref{fig2}.}%
\label{fig3a}%
\end{figure}\begin{figure}[th]
\centering
\includegraphics[width=3.in]{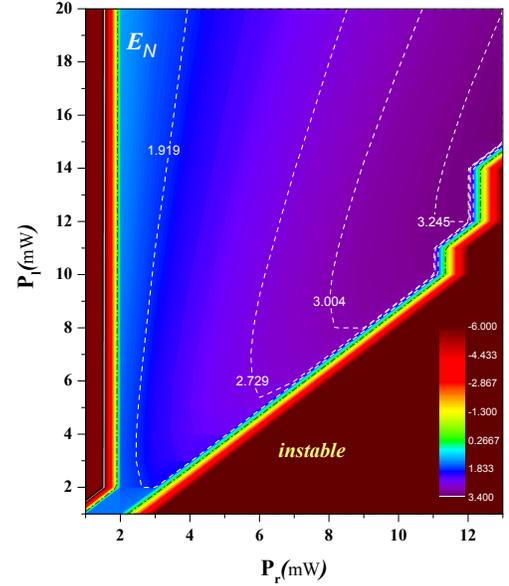}
 \vspace{-1.0cm}
 \caption{ (Color online) Logarithmic
negativity $E_{N}$ between the optical output modes of the two subcavities
versus the input powers $P_{l}$ and $P_{r}$ for $\Omega_{r}=-\Omega_{l}=-\Omega_{m}$. Again the other parameters are the same as those in
Fig.~\ref{fig2}.}%
\label{fig3b}%
\end{figure}In Fig. \ref{fig2} we have plotted the logarithmic negativity
$E_{N}$ versus the normalized frequency of the output cavity mode $\Omega
_{l}/\Omega_{m}$ for two different values of membrane quality factor
$Q=10^{4}$ and $Q=15\times10^{4}$. We have assumed an experimental situation
\cite{thompson,teuf} representing a membrane with vibrational frequency
$\Omega_{m}/2\pi=10$MHz and mass $m=10$ng. The right side subcavity has
damping rate $\kappa_{r}=0.4\Omega_{m}$ and the laser power imping on this
subcavity is assumed to be $P_{r}=10$mW. The left side subcavity damping rate
is $\kappa_{l}=0.1\Omega_{m}$ with the laser power $P_{l}=48$mW. The
temperature of membrane's reservoir is $T=1$K and the subcavities' detunings
have been fixed at $\Delta_{r}=-\Delta_{r}=-\Omega_{m}$ with $\Omega
_{r}=-\Omega_{m}$. Figure (\ref{fig2}) shows that the entanglement between
output cavity fields is maximum around $\Omega_{l}=\Omega_{m}$. Also we see
that by increasing the quality factor of the mechanical resonator one can
increase the entanglement between the subcavities' output fields.

A more interesting situation is depicted in Fig.~\ref{fig3a} which shows how
entanglement between the output cavity fields depends on the subcavities
damping rates. This figure indicates that the entanglement reaches its maximum
around $\kappa_{r}\sim0.35\Omega_{m}$ and $\kappa_{l}\sim0.2\Omega_{m}$ and
out of this region entanglement quickly decreases. Note that the maximum of
entanglement is approached around small values of damping rates which is close
to the instability threshold. Whereas, Fig.~\ref{fig3b} shows proper values of
input powers which maximize entanglement. This figure reveals that by
increasing the input powers one can definitely improve the entanglement
between the output modes, even though at the same time the instability region
is extended. Finally, we note that the generated CV entanglement can be
verified from the measurement record by applying a generalized version of
Duan's inequality~\cite{wool}.

\section{Dense coding with an optomechanical source}

So far we have shown that an optomechanical cavity in the form of a
membrane-in-the-middle geometry can be used to generate entanglement between
two optical fields. Now we show that these optical modes are sufficiently
entangled to be used for implementing the protocol of dense
coding.\begin{figure}[th]
\centering
\includegraphics[width=2.7in]{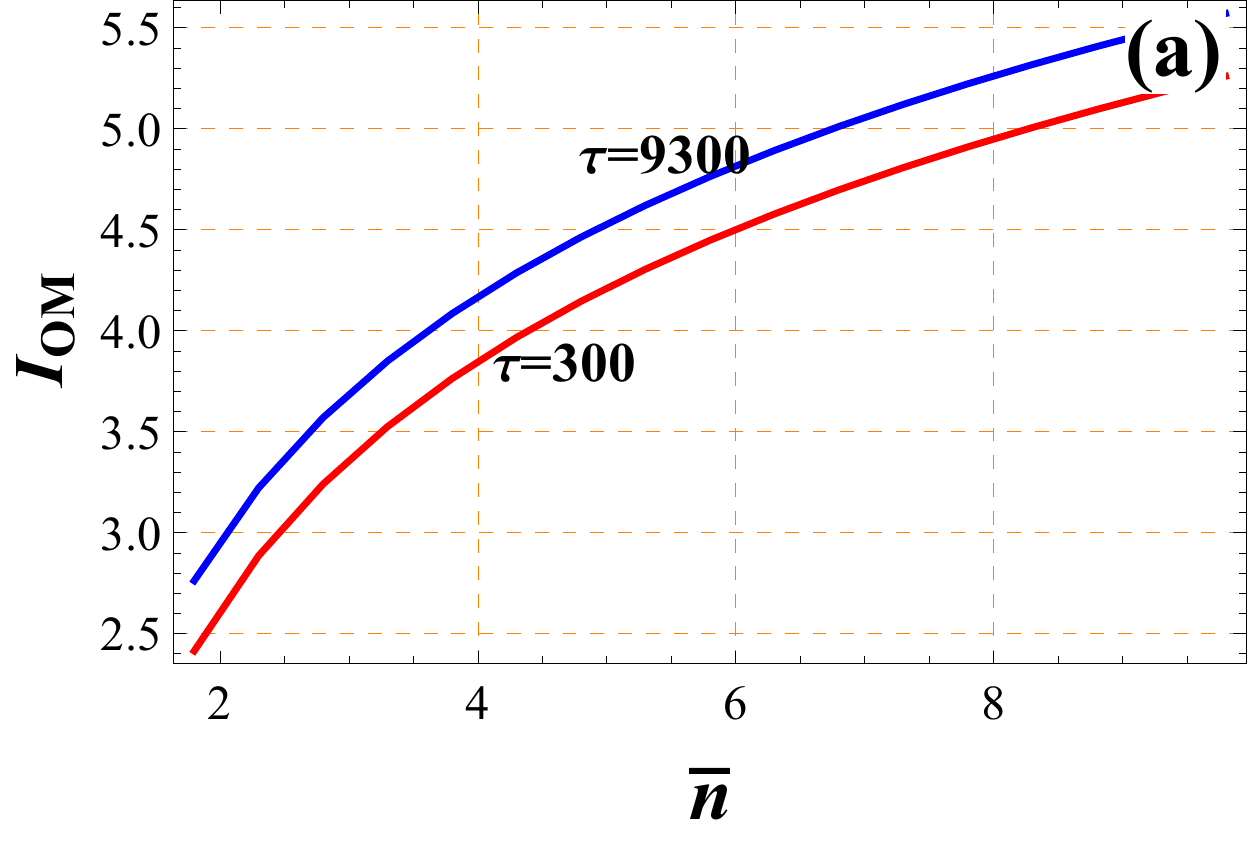}
\includegraphics[width=2.7in]{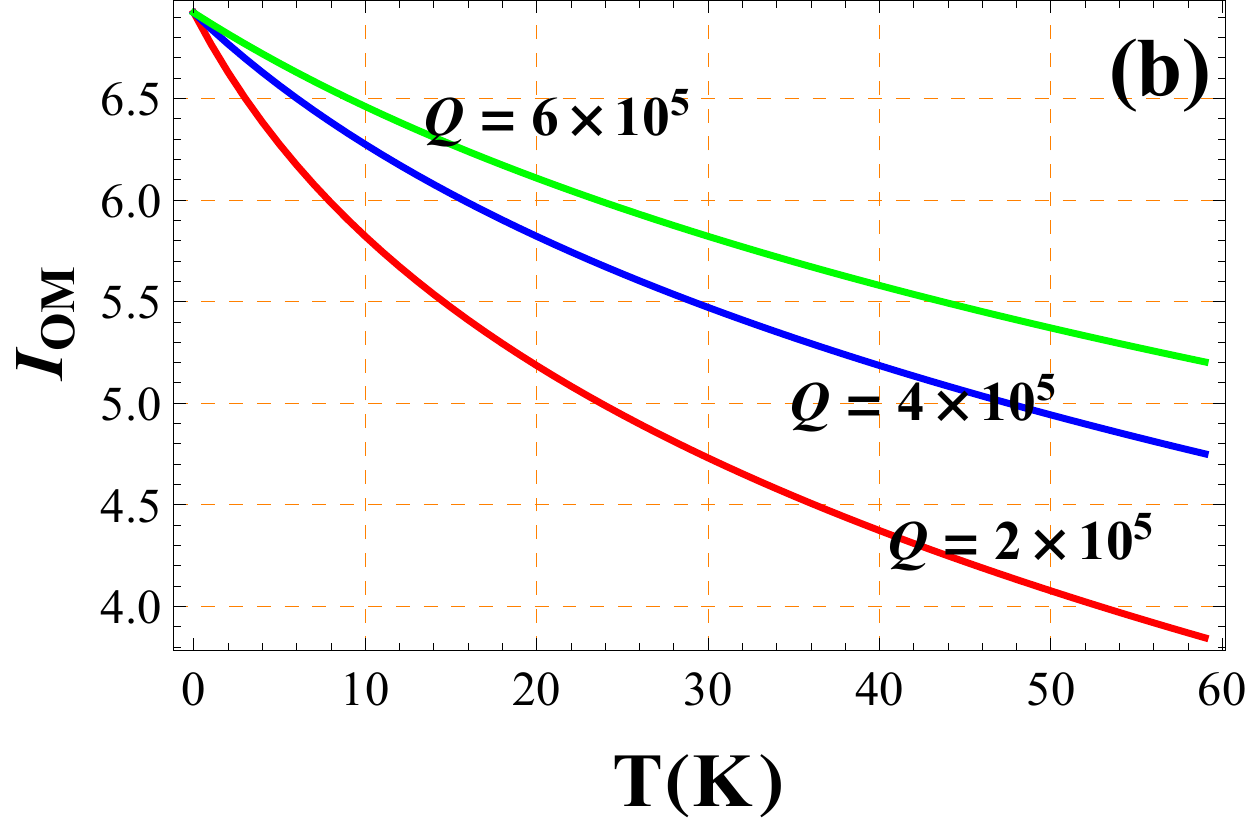}
\vspace{-0.25cm}
\caption{ (Color online)
Optomechanical dense coding rate $I_{OM}$ (a) versus the average photon
numbers for two different values of the laser bandwidth and (b) versus the
temperature of the cavity for three different values of the mirror's quality
factor. Here we consider $\Omega_{r}=-\Omega_{l}=-\Omega_{m}$ and the other
parameters are the same as in Fig.~\ref{fig2}.}
\label{fig4}
\end{figure}

The scheme is sketched in Fig.~\ref{fig1}, where the entangled optical beams
at the output of the cavity are labelled by $l$ and $r$, with mode $i=l,r$
having quadrature fluctuations $\delta X_{i}^{out}$ and $\delta Y_{i}^{out}$.
The left mode $l$ is sent to Alice, while the right mode $r$ is sent to Bob.
Then, Alice encodes a Gaussian complex signal $\alpha_{s}=(X_{s}+iY_{s}%
)/\sqrt{2}$ by applying the displacement operator $D(\alpha_{s})$ on mode $l$
(with $V_{s}$ being the variance of each real Gaussian variable $X_{s}$ and
$Y_{s}$). The output mode, with quadrature fluctuations $\delta X_{l}%
^{out}+X_{s}$ and $\delta Y_{l}^{out}+Y_{s}$, is sent to Bob through a
noiseless quantum channel. At his station, Bob combines the incoming signal
mode with mode $r$ in a balanced beam splitter, of which he homodynes the two
output ports, measuring the position fluctuation of \textquotedblleft%
$+$\textquotedblright\ and the momentum fluctuaction of \textquotedblleft%
$-$\textquotedblright. In other words, Bob detects the two operators
\begin{align}
\delta X_{+}  &  =\frac{1}{\sqrt{2}}\left(  \delta X_{l}^{out}+X_{s}+\delta
X_{r}^{out}\right)  ~,\nonumber\\
\delta Y_{-} &  =\frac{1}{\sqrt{2}}\left(  \delta Y_{l}^{out}+Y_{s}-\delta Y_{l}^{out}\right) ~. \label{quad}
\end{align}
One can easily check that these operators have the same variance, i.e.,
\begin{equation}
\langle\delta X_{+}^{2}\rangle=\langle\delta Y_{-}^{2}\rangle=\frac{1}{2}(L+R-2C+V_{s}):=V_{B}~. \label{quad0}
\end{equation}
It is also easy to compute the conditional entropy $V_{A|B}$ which quantifies
the remaining entropies of $X_{s}$ and $Y_{s}$ given Bob's homodyne
detections. This is given by
\begin{align}
V_{A|B}  &  =\langle X_{s}^{2}\rangle-\frac{\langle X_{s}(\delta X_{+}
)\rangle^{2}}{\langle\delta X_{+}^{2}\rangle}\label{condENT}\\
&  =\langle Y_{s}^{2}\rangle-\frac{\langle Y_{s}(\delta Y_{-})\rangle^{2}
}{\langle\delta Y_{-}^{2}\rangle}=V_{s}-\frac{V_{s}^{2}}{2V_{B}}~.\nonumber
\end{align}
\begin{figure}[th]
\centering
 \vspace{-1.0cm}
\includegraphics[width=3.2in]{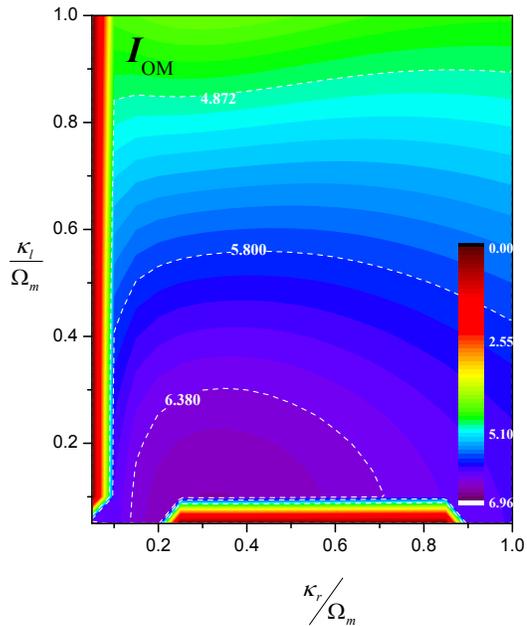}
 \vspace{-1.0cm}
 \caption{ (Color online) Optomechanical
dense coding rate $I_{OM}$ versus the normalized damping rates $\kappa
_{l}/\Omega_{m}$ and $\kappa_{r}/\Omega_{m}$ at a fixed temperature $T=1$ K
and $\Omega_{r}=-\Omega_{l}=-\Omega_{m}$. The other parameters are the same as
in Fig.~\ref{fig2}.}%
\label{fig5}%
\end{figure}

Now, using Eqs.~(\ref{quad0}) and~(\ref{condENT}), we can compute Alice and
Bob's mutual information
\begin{align}
I(A  &  :B)=\log_{2}\frac{V_{s}}{V_{A|B}}\label{mutual}\\
&  =\log_{2}\left(  1+\frac{V_{s}}{L+R+2C}\right)  \mathrm{~}.\nonumber
\end{align}
Here the signal power can be written as $V_{s}=\bar{n}_{s}+1/2$. In turn, the
thermal number $\bar{n}_{s}$ can be written in terms of the mean number of
photons $\bar{n}$ which are sent to Bob through the noiseless quantum channel.
This mean photon number represents the energetic constraint of the protocol,
and is equal to $\bar{n}=\bar{n}_{l}^{out}+\bar{n}_{s}$, where $\bar{n}%
_{l}^{out}=\langle(b_{l}^{out})^{\dagger}b_{l}^{out}\rangle$ is the average
number of photons in the output cavity mode $l$. Then, we can write the signal
power as $V_{s}=(\bar{n}+1)-L$. Finally, by replacing $V_{s}$ in
Eq.~(\ref{mutual}) we get the mutual information of Alice and Bob $I(A:B)$ in
terms of the energetic constraint $\bar{n}$. This quantity represents the
dense coding rate $I_{OM}(\bar{n})$ which is achievable by using our
optomechanical source.\begin{figure}[th]
\centering
 \vspace{-0.5cm}
\includegraphics[width=3.2in]{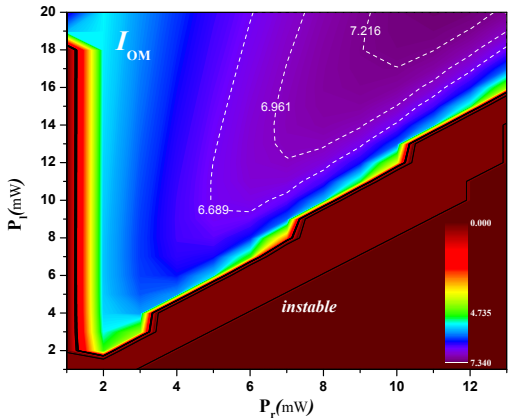}
 \vspace{-0.5cm}
 \caption{ (Color online) Optomechanical
dense coding rate $I_{OM}$ versus the input powers $P_{l}$ and $P_{r}$ at a
fixed temperature $T=1$~K and $\Omega_{r}=-\Omega_{l}=-\Omega_{m}$. The other
parameters are the same as Fig.~\ref{fig2}.}%
\label{fig6}%
\end{figure}

In Fig.~\ref{fig4}~(a) we have plotted the optomechanical dense coding rate
$I_{OM}$ in terms of the energetic constraint $\bar{n}$ for two different
values of laser bandwidth $\tau$. As expected, $I_{OM}$ is increasing in
$\bar{n}$, and also in the bandwidth $\tau$. Then, as shown in Fig.~\ref{fig4}~(b), we see that $I_{OM}$ is relatively robust with respect to the temperature
of the cavity. The dependence of the optomechanical rate\ on the cavity
dampings is illustrated in Fig.~\ref{fig5}, where we can see that $I_{OM}$ is
maximum around $\kappa_{r}\simeq0.25\Omega_{m}$ and $\kappa_{l}\simeq
0.1\Omega_{m}$. Finally, in Fig.~\ref{fig6} we also show the influence of the
input powers.\begin{figure}[th]
\centering
\includegraphics[width=3.2in]{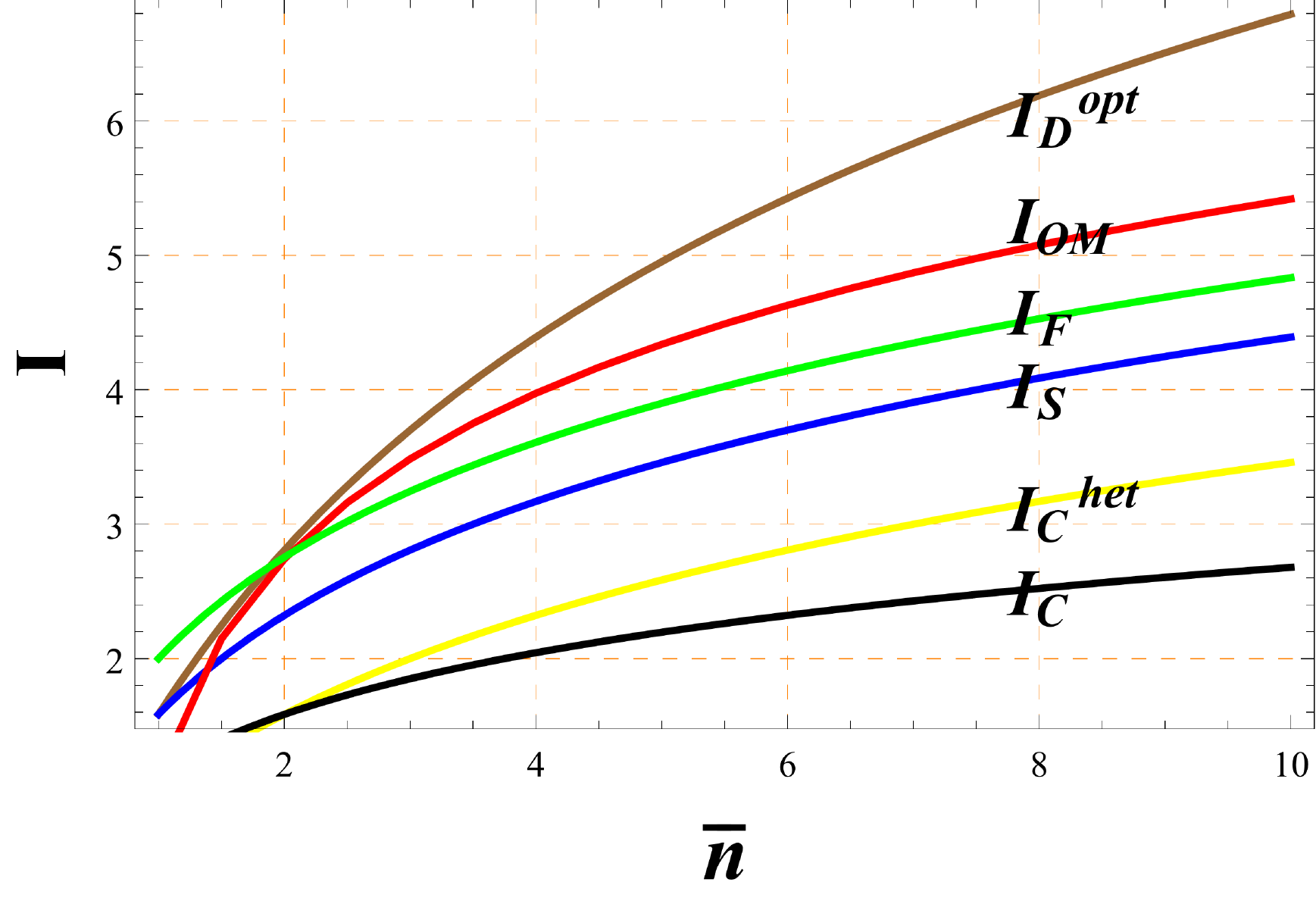}\caption{(Color online) The
optomechanical dense coding rate $I_{OM}$ is plotted in terms of the photon
number $\bar{n}$, and compared with the dense coding capacity $I_{D}^{opt}$,
the classical capacity $I_{F}$\ (Fock states and photon counting), the rate
$I_{S}$\ achievable by squeezed states and homodyne, the rate $I_{C}^{het}%
$\ achievable by coherent states and heterodyne, and finally, the rate $I_{C}$
reachable by coherent states and homodyne. We consider $T=1$~K and $\Omega
_{l}=-\Omega_{r}=\Omega_{m}$. The subcavities' detunings are $\Delta
_{l}=-\Delta_{r}=\Omega_{m}$, while the other optomechanical parameters are
$\Omega_{m}/2\pi=10$~MHz, $\kappa_{r}=0.4\Omega_{m}$, $P_{r}=10$ mW, $L=1$ mm,
$\kappa_{l}=0.1\Omega_{m}$, $P_{l}=48~$mW, $m=10~$ng, and $Q=15\times10^{4}$.}%
\label{fig7}%
\end{figure}

As shown in Fig.~\ref{fig7}, the optomechanical dense coding rate
$I_{OM}(\bar{n})$ is able to approximate the dense coding capacity~\cite{bra}
$I_{D}^{opt}(\bar{n})=\mathrm{log}_{2}(1+\bar{n}+\bar{n}^{2})$ for low photon
numbers $\bar{n}\simeq2$, remaining suboptimal at higher energies. As we show
in the same figure, for $\bar{n}>2$ the optomechanical dense coding rate
$I_{OM}(\bar{n})$ outperforms all the rates associated with one-way quantum
communication from Alice to Bob at the same energy. Indeed, it beats the
classical capacity of the noiseless quantum channel $I_{F}(\bar{n})=(1+\bar
{n})\mathrm{log}_{2}(1+\bar{n})-\bar{n}\mathrm{log}_{2}\bar{n}$, which can be
reached by encoding in Fock states and decoding by photon
counting~\cite{Caves,Yuen}. Then, it clearly outperforms the rate $I_{S}%
(\bar{n})=\mathrm{log}_{2}(1+2\bar{n})$ achievable by squeezed states and
homodyne detection~ \cite{Yama}, the rate $I_{C}^{het}(\bar{n})=\mathrm{log}%
_{2}(1+\bar{n})$ reachable by coherent states and heterodyne
detection~\cite{Yama,She}, and finally the rate $I_{C}(\bar{n})=\mathrm{log}%
_{2}(\sqrt{1+4\bar{n}})$ which can be reached by coherent states and homodyne
detection~\cite{bra,ralph}.

\section{Conclusion}

In this paper, we have shown that continuous-variable dense coding can be implemented using an
optomechanical cavity as the source of the entanglement. We have considered a
high-finesse cavity with a membrane-in-the-middle geometry, i.e., formed by
two fixed end mirrors and a perfectly-reflecting movable mirror in the middle.
The dynamics of the system has been investigated by solving the quantum
Langevin equations. After their linearization, we have analyzed the stationary
entanglement which can be established between the two output optical modes of
the cavity, showing its behaviour in terms of the main optomechanical
parameters, such as the mechanical damping rates or the laser input powers.

Using the optical entanglement generated by the cavity we have then
implemented the protocol of continuous-variables dense coding. We have computed the
optomechanical dense coding rate, studying its behavior in terms of the
various system parameters, including the input powers, the damping rates, and
the quality factor, mass and temperature of the movable mirror. We have shown
how this rate approximates the dense coding capacity at low photon numbers
($\bar{n}\simeq2$), and outperforms the one-way classical capacity of the
noiseless quantum channel at higher energies ($\bar{n}>2$). As a result, we
have proven how an optomechanical cavity is able to generate an amount of
optical entanglement which is strong enough to implement a standard protocol
of quantum information.

\subsection*{Acknowledgements}

We thank M. Woolley for helpful comments. C.W. is supported by NSERC. The work
of S.P. has been supported by EPSRC and the Leverhulme Trust. S.B. is grateful for support from the Alexander von
Humboldt foundation.

\end{document}